\documentclass{article}

 \usepackage[main, final,nonatbib]{neurips_2025}

\usepackage[utf8]{inputenc} 
\usepackage[T1]{fontenc}    
\usepackage{hyperref}       
\usepackage{url}            
\usepackage{booktabs}       
\usepackage{amsfonts}       
\usepackage{nicefrac}       
\usepackage{microtype}      
\usepackage{xcolor}         
\usepackage{amsmath}
\usepackage{graphicx}
\usepackage{wrapfig}
\usepackage{cprotect} 
\usepackage{caption}
 \usepackage[backend=biber,sorting=none,style=numeric]{biblatex}
\title{Searching solo for the invisible at the Compact Muon Solenoid}

\author{%
  Abhishikth Mallampalli \\
  For the Compact Muon Solenoid (CMS) Collaboration\\
}

\usepackage[utf8]{inputenc}
\begin{document}

\begin{titlepage}
    \begin{center}
        \includegraphics[width=\textwidth]{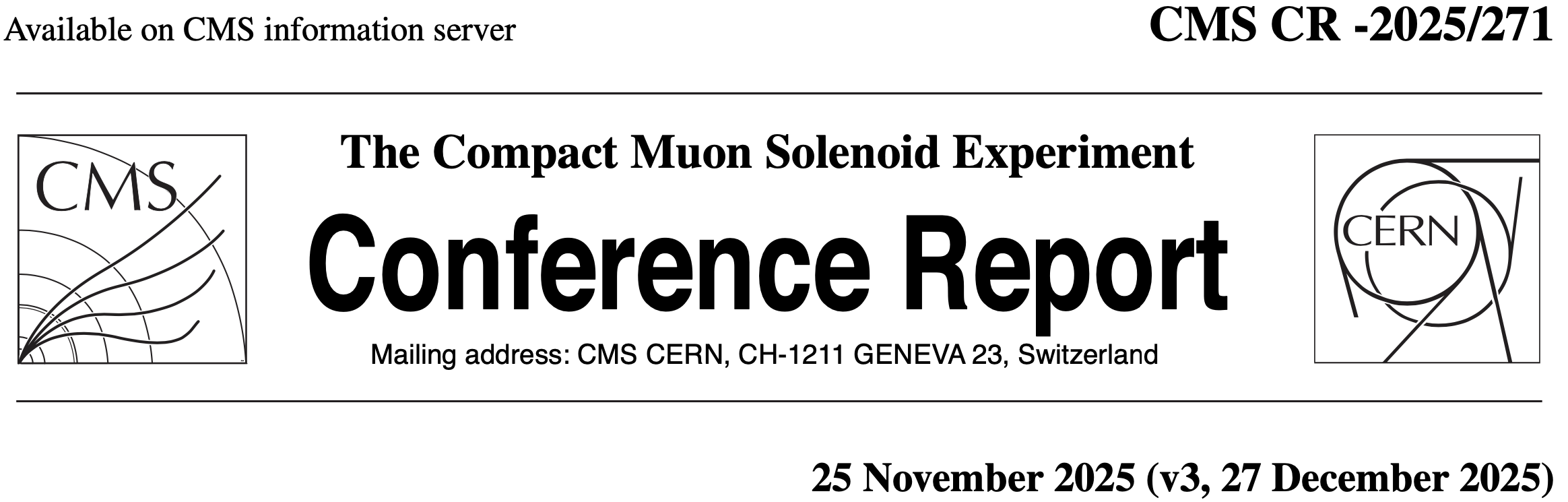} 
    
        \vspace{2cm}

        {\Huge \bfseries Searching solo for the invisible at CMS \par}
        
        \vspace{1.5cm}

        {
        \sffamily
        {\LARGE \textbf{Abhishikth Mallampalli}} \hspace{0.02cm} 
{\large for the CMS Collaboration}
}
        
        \vspace{2cm}

        \begin{abstract}
        Despite the success of the Standard Model (SM), several fundamental questions remain unanswered, such as the nature of dark matter (DM), motivating searches for new physics. This paper summarizes three recent searches for new physics in proton-proton collisions at a center-of-mass energy of $\sqrt{s}=13$~TeV, using data recorded with the Compact Muon Solenoid (CMS) detector at the CERN Large Hadron Collider (LHC). The searches focus on “mono-X” final states, characterized by a large imbalance in transverse momentum recoiling against a single visible SM particle ($X$), and serve as powerful probes of new physics scenarios. Results are presented for searches in the pencil-jet (low-multiplicity jet), mono-photon, and mono-top final states, using CMS Run 2 data corresponding to an integrated luminosity of 138~fb$^{-1}$. No significant excess of events beyond SM predictions is observed, and the results are used to set stringent exclusion limits on various new physics scenarios, including simplified DM models and models of large extra spacetime dimensions.
        \end{abstract}

        \vspace{2cm}

        {Presented at LP2025 32nd International Symposium on Lepton Photon Interactions at High Energies}

    \end{center}
\end{titlepage}


\section{Introduction}
Determining the nature of dark matter (DM) is one of the major open questions of particle physics and cosmology. While astrophysical observations provide compelling evidence for its existence, its particle nature is unknown. The LHC provides a powerful opportunity to probe models of dark matter by analyzing the rich data produced in high-energy proton-proton (pp) collisions. In pp collisions, the initial momentum in the plane transverse to the beam direction is essentially zero. Due to the conservation of momentum and the nearly hermetic nature of the CMS detector, the vector sum of the transverse momenta of all reconstructed final-state particles is expected to vanish, within measurement uncertainties and after accounting for known Standard Model (SM) particles such as neutrinos. If a significant imbalance in transverse momentum is observed (referred to as missing transverse momentum --- $p_{\mathrm{T}}^\text{miss}$), it implies the presence of one or more particles that have escaped detection --- which could be signals of new physics.
\noindent\begin{minipage}[t]{0.47\textwidth}
  \vspace{0pt}
  A well-motivated signature for DM production is the "mono-X" topology, where a pair of DM particles ($\chi\bar{\chi}$) is produced in association with a single visible SM particle ($X$). The three recent searches from CMS presented here look for this signature: a large ($> 200$ GeV) $p_{\mathrm{T}}^\text{miss}$ imbalance and one "solo" SM object (a jet, photon, or top quark), hence the title "Searching solo for the invisible". Fig.~\ref{fig:event_display} shows one such mono-X event, recorded by the CMS detector on July 16, 2017. Individual events with this topology can arise from various sources, including dark matter production, extra spatial dimensions, or simply SM backgrounds. Consequently, a detailed statistical analysis of the full dataset is required to distinguish a potential new physics signal from these background processes.
\end{minipage}%
\hfill
\begin{minipage}[t]{0.5\textwidth}
  \vspace{0pt}
  \includegraphics[width=\linewidth]{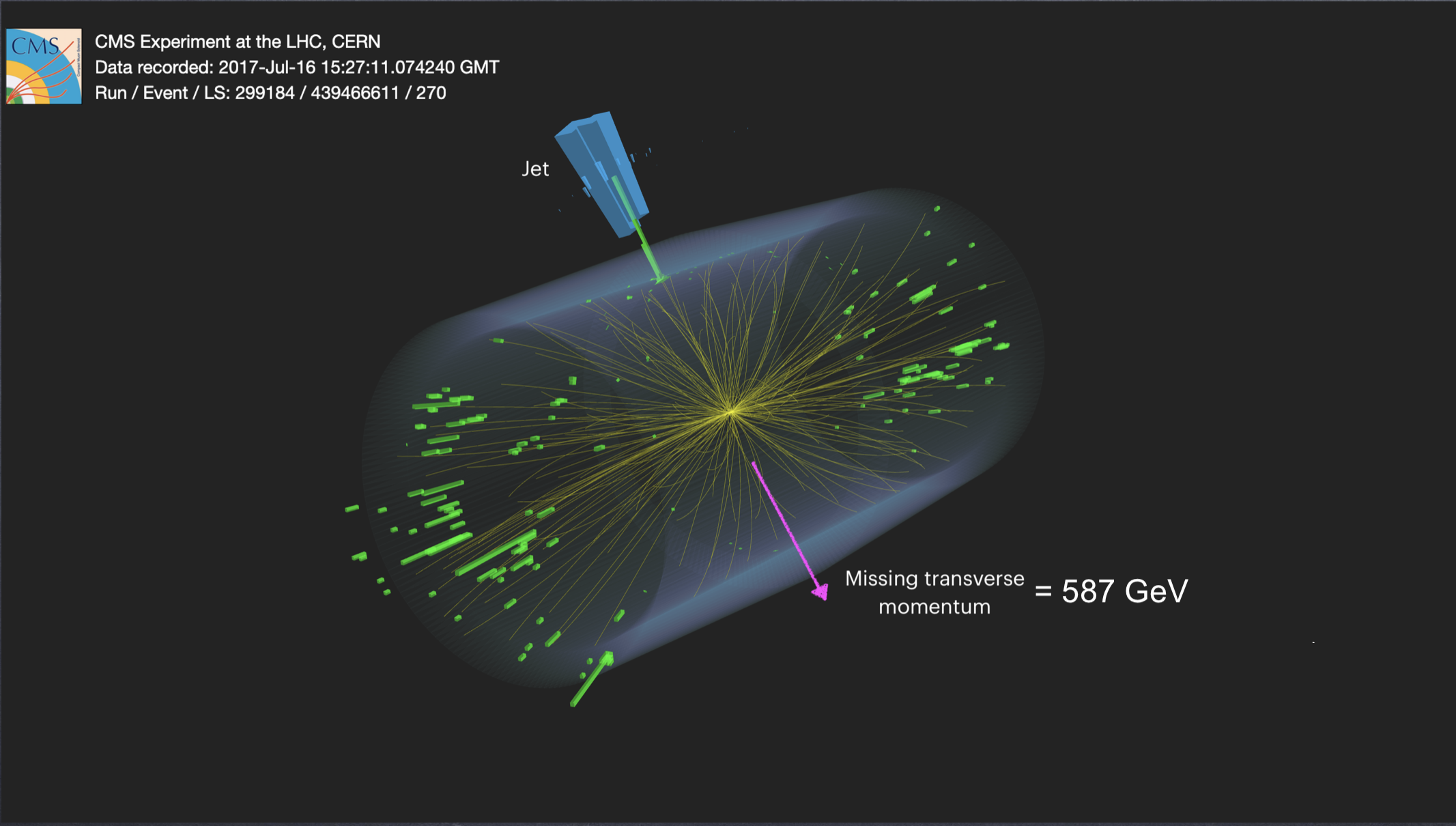}
  \captionof{figure}{An example of a mono-X event, recorded by the CMS detector.}
  \label{fig:event_display}
\end{minipage}


The searches described in this paper have been performed at a center-of-mass energy of $\sqrt{s}=13$~TeV using the Run 2 data collected by the CMS detector, corresponding to a total integrated luminosity of 138~fb$^{-1}$. The detailed description of the CMS detector is available in ~\cite{CMS:2008xjf}. The following sections outline each of the three searches. 

\section{Pencil-Jet Search}
This search~\cite{CMS-PAS-SUS-23-017} focused on a low-multiplicity pencil-thin jet signature and is the first search for this signal at the LHC.
\noindent\begin{minipage}[t]{0.47\textwidth}
  \vspace{0pt}
   The analysis targets a simplified model~\cite{ABERCROMBIE2020100371} as shown by the Feynman diagram in Fig.~\ref{fig:feynmandiag}. The primary process involves DM pair production, followed by the radiation of a dark Z' boson from one of the DM particles, as illustrated in Fig.~\ref{fig:feynmandiag}. The Z' boson, emitted via final-state radiation (FSR), can decay back into SM particles, producing a visible signature, while the DM particles themselves contribute to significant missing transverse momentum. The focus is on scenarios where the Z' boson predominantly decays into quarks, and is expected to be reconstructed in the detector as a jet with very narrow cone of radiation, called the pencil-jet.

  \vspace{0.2cm}

  Supervised machine learning (ML) and data augmentation techniques are used to enhance the sensitivity to this rare signal against the dominant Z($\nu\nu$)+jets and W($l\nu$)+jets backgrounds and improve robustness to input feature mismeasurements. Unlike traditional mono-X searches, the signal being studied has a significant contribution from final state radiation (FSR) in addition to initial state radiation (ISR) as described in Ref.~\cite{MonoZprime_Theory}.
\end{minipage}%
\hfill
\begin{minipage}[t]{0.5\textwidth}
  \vspace{0pt}
  \includegraphics[width=0.9\linewidth]{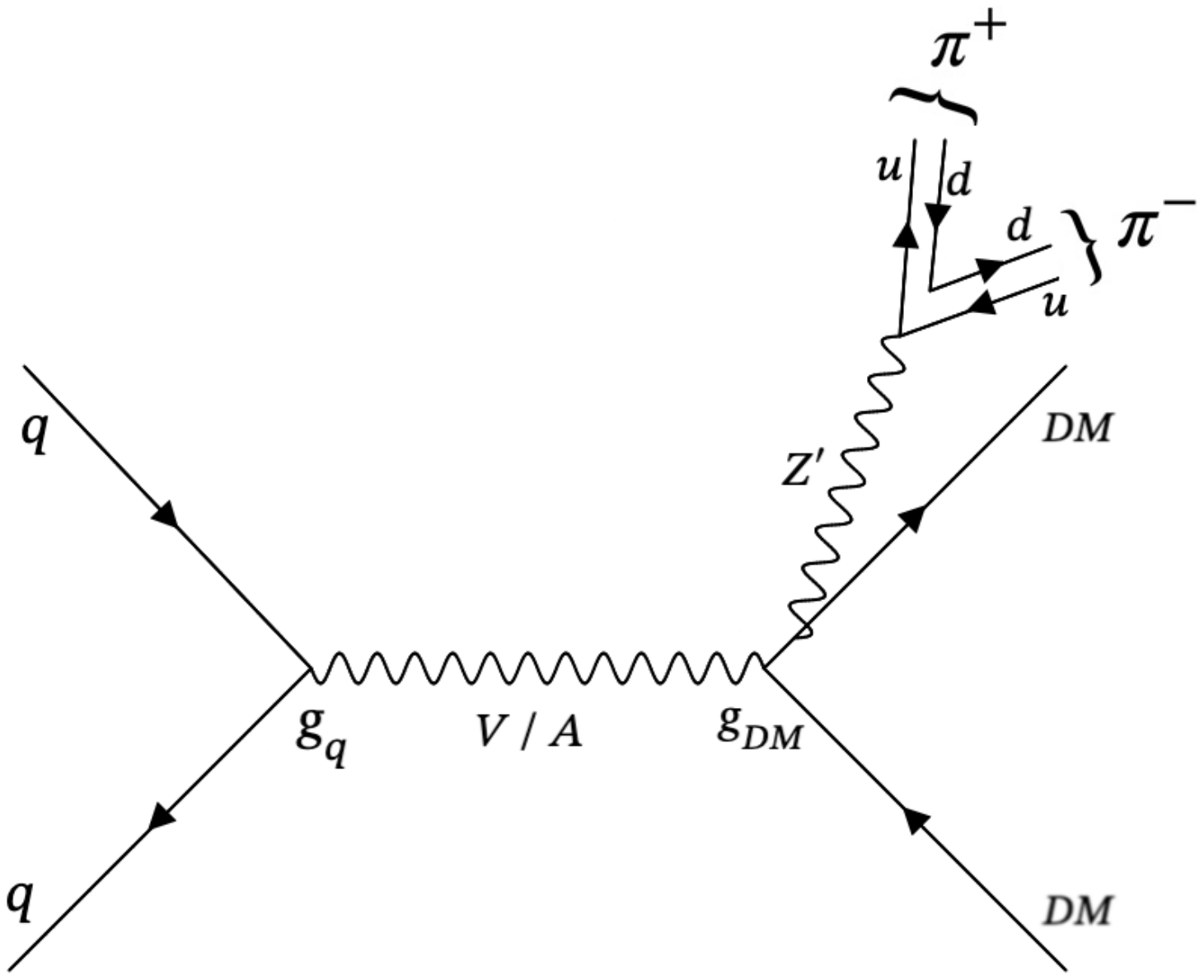}
  \captionof{figure}{Representative leading order Feynman diagram of the signal process, where $\chi$ is the DM candidate, V/A represents the vector or axial vector mediator, $g_q$ and $g_{DM}$ represents the mediator coupling to the quarks and DM respectively. Due to the low mass of the Z' boson ($\approx 1 ~$ GeV), it primarily decays to a low multiplicity final state, typically two pions.}
  \label{fig:feynmandiag}
\end{minipage}


The free parameters in the model are the DM particle mass ($100$ GeV--$1$ TeV), mediator mass (TeV scale) and Z' mass ($\sim 1$ GeV). The couplings are fixed based on recommendation of the ATLAS/CMS-DM forum~\cite{ABERCROMBIE2020100371}. 

\begin{figure}[htb!]
    \centering
        \includegraphics[width=0.8\columnwidth]{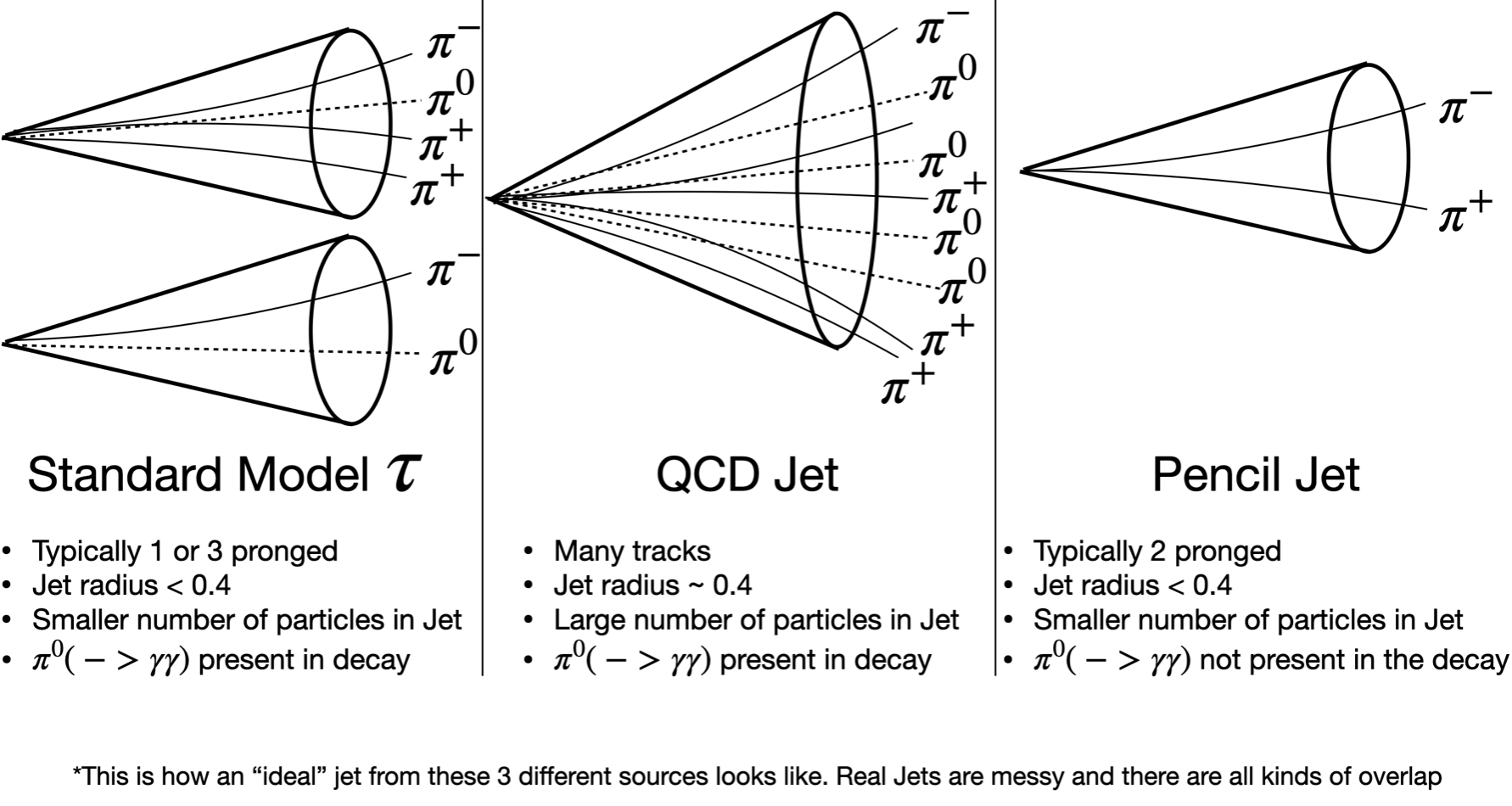}
        \caption{Distinguishing features of a pencil-jet against the standard hadronic $\tau$ and QCD jets.}
        \label{fig:penciljet}
\end{figure}

Fig.~\ref{fig:penciljet} outlines the distinguishing features of a pencil-jet compared to the SM hadronic $\tau$ decays and Quantum Chromodynamic (QCD) jets.

To access jet substructure information, each pencil-jet is matched to a jet from a collection of jets clustered using the infrared- and collinear-safe anti-\ensuremath{k_{\mathrm{T}}}~algorithm~\cite{Cacciari:2008gp, Cacciari:2011ma} with a distance parameter of 0.4 (AK4 jets). This matched jet is referred to as the associated jet.

The main selections include $p_{\mathrm{T}}^\text{miss}$ filters (which remove events with known spurious contributions to the $p_{\mathrm{T}}^\text{miss}$), azimuthal angle ($\phi$) between $p_{\mathrm{T}}^\text{miss}$ direction and the leading pencil-jet to be larger than 0.5 radians (this requirement ensures that the pencil-jet and $p_{\mathrm{T}}^\text{miss}$ are back-to-back), $p_{\mathrm{T}}^\text{miss}$ $>250$ GeV, associated jet cleaning criteria, and selections to reject events having electrons, muons, and photons.

Main uncertainties include statistical uncertainties, higher order QCD corrections to the dominant Z($\nu\nu$)+jets and W($l\nu$)+jets backgrounds and pencil-jet reconstruction uncertainties. The pencil-jet reconstruction performance and uncertainties are derived by modifying the SM $\tau$ object definitions. 

\begin{figure}[htb!]
    \centering
        \includegraphics[width=0.8\columnwidth]{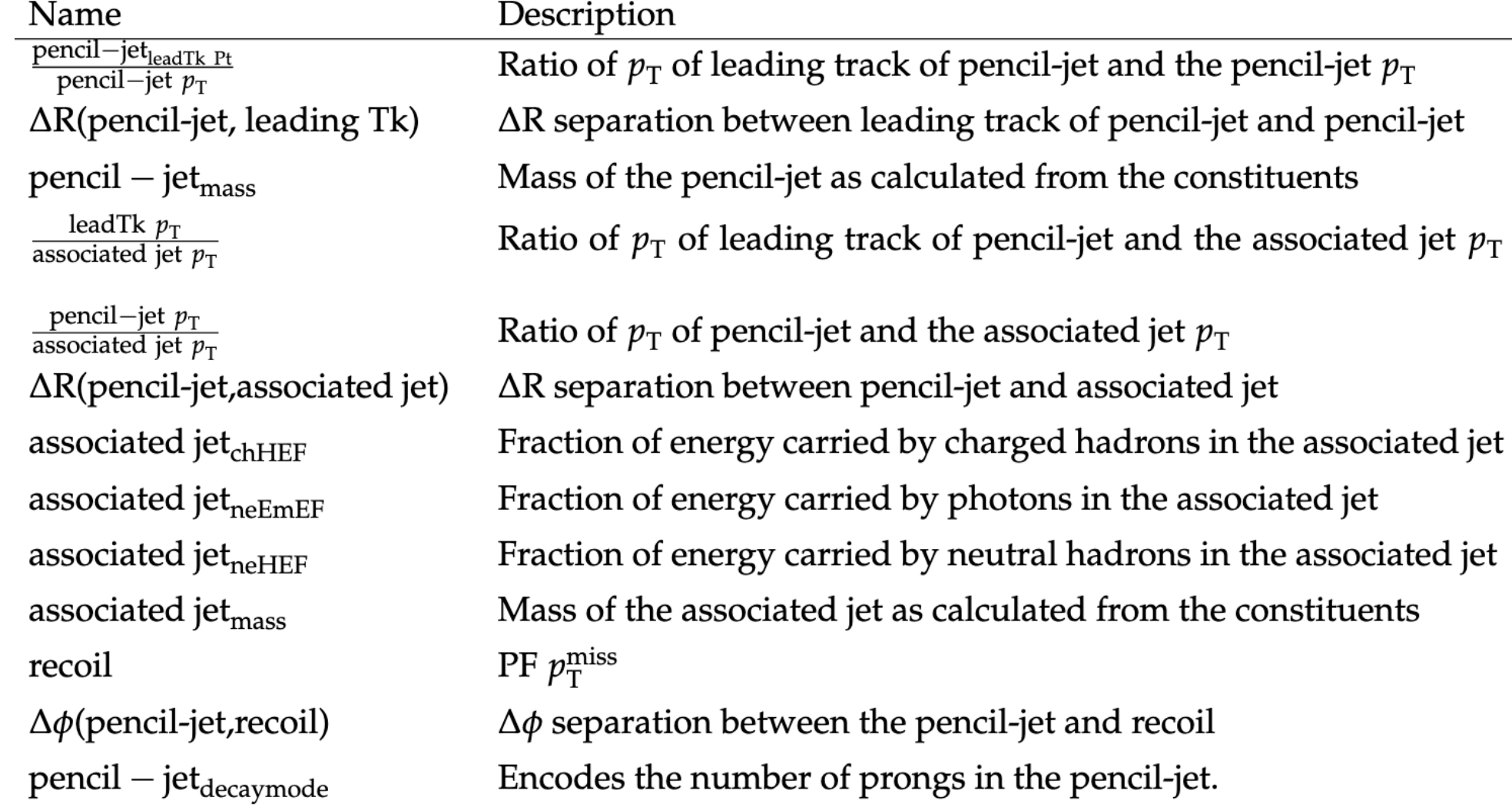}
        \caption{Distinguishing features of a pencil-jet against the standard hadronic $\tau$ and QCD jets.}
        \label{fig:input_features}
\end{figure}

A combination of a neural-network and gradient-boosted decision tree is used to construct a single discriminating variable out of a combination of 13 physically-motivated input features, defined in Fig.~\ref{fig:input_features}. Four control samples (single-electron, double-electron, single-muon, double-muon) are defined for each era of data processing by reversing the rejection criteria of electrons and muons, and they are used to validate the performance of the ML model and data-simulation scale factors. This discriminating variable is then used to extract the signal using a maximum likelihood fit.

\begin{figure}[htb!]
    \centering
        \includegraphics[width=1.0\columnwidth]{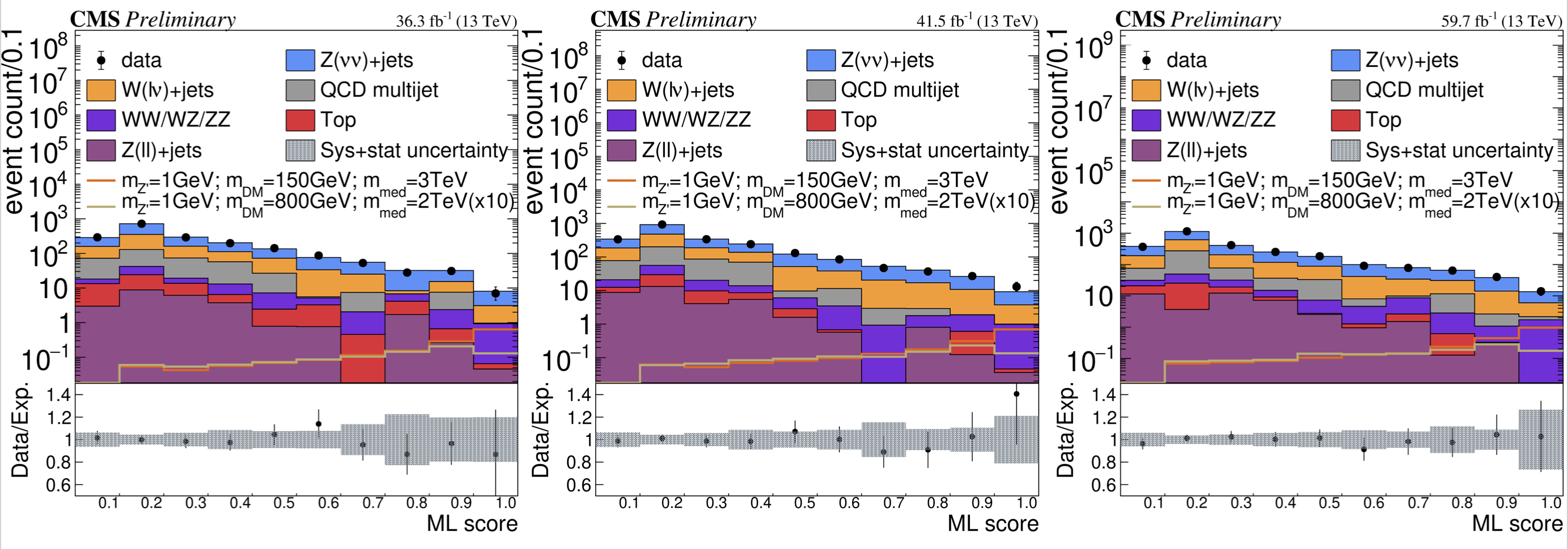}
        \caption{Postfit plots showing the data-simulation agreement for 2016 (left), 2017 (middle), and 2018 (right).}
        \label{fig:zprime_results}
\end{figure}

Fig.~\ref{fig:zprime_results} shows the postfit plots for the 3 years of data-collection in Run 2. No excess of events over the SM background expectation is observed. 

\begin{figure}[ht!]
    \centering
        \includegraphics[width=1.0\columnwidth]{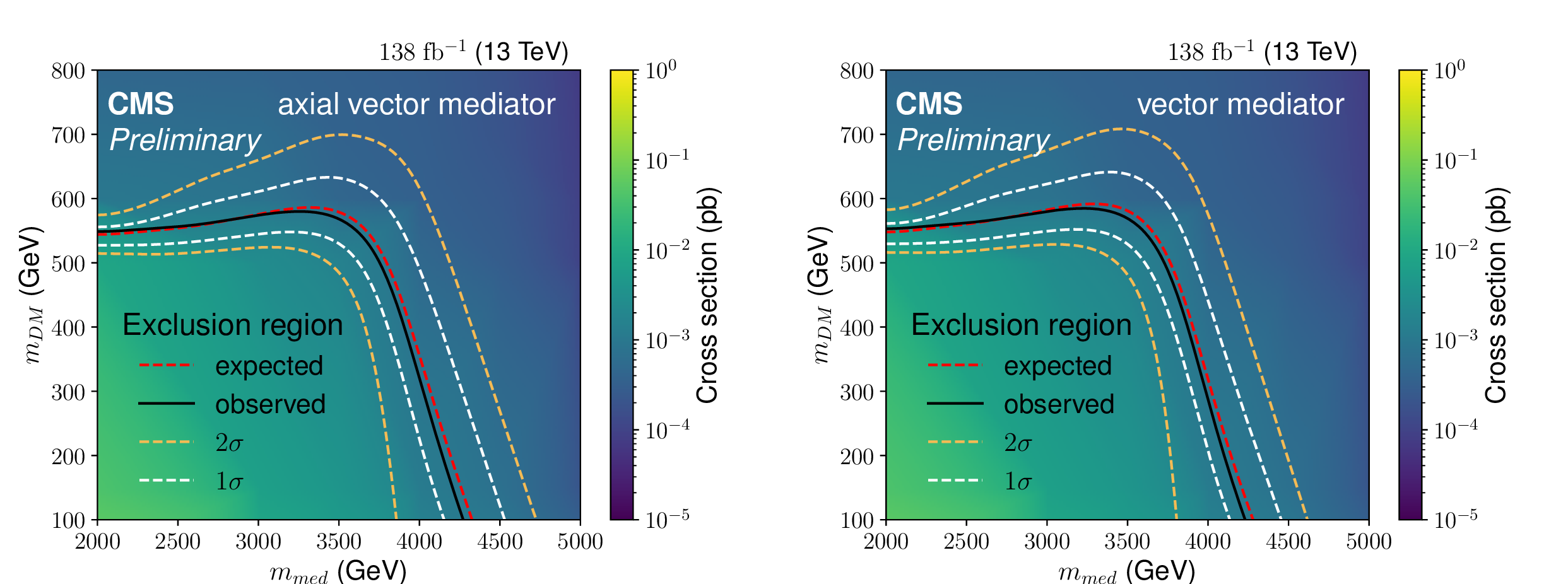}
        \caption{Expected and observed exclusion region in the $m_{\mathrm{DM}}-m_{\mathrm{med}}$ plane for the axial vector mediator (left) and vector mediator (right), for $m_{Z'}= 1 $ GeV. The parameter space to the lower-left in both the panels is excluded at the 95\% CL level. The color scale represents the theoretical cross section of the signal process.}
        \label{fig:zprime_interpretation}
\end{figure}

The results are interpreted in the context of simplified models with vector and axial-vector mediators, and 95\% confidence level (CL) upper limits are set on DM production cross section, as shown in Fig.~\ref{fig:zprime_interpretation}. Mediator masses up to $\sim 4250$ GeV are excluded for a DM mass of $\sim 100$ GeV, and up to $\sim 3500$ GeV for a DM mass of $\sim 550$ GeV.

\section{Mono-Photon Search}

\begin{figure}[ht!]
    \centering
        \includegraphics[width=0.8\columnwidth]{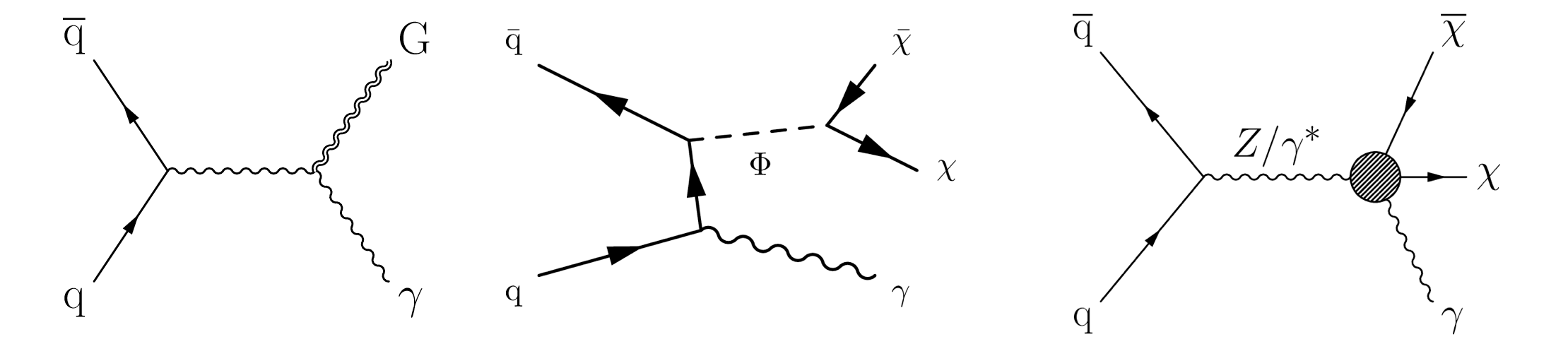}
        \caption{Leading-order diagrams of graviton (G) production in the ADD model (left), simplified DM model (center), and EW-DM effective interaction (right), with a final state comprising a photon and large $p_{\mathrm{T}}^\text{miss}$. $\chi$ is the DM candidate, and $\Phi$ in the simplified DM model represents a vector or axial-vector mediator.}
        \label{fig:monophoton_feyndiag}
\end{figure}

This analysis~\cite{cmscollaboration2025searchnewphysicsfinal} targets a signature where there is a SM photon produced along with a large $p_{\mathrm{T}}^\text{miss}$. This signature can be produced by DM scenarios and extra spacetime dimensions~\cite{Arkani_Hamed_1998,Arkani_Hamed_1999,Arkani_Hamed_2000} as shown in Fig.~\ref{fig:monophoton_feyndiag}, in addition to the dominant SM backgrounds of Z($\nu\nu$)+$\gamma$ and W($l\nu$)+$\gamma$. In addition, stray electromagnetic calorimeter (ECAL) clusters produced by mechanisms other than pp collisions can be misidentified
as photons. Beam halo muons in particular, which accompany the proton beams and traverse the detector longitudinally, contribute significantly to the rate of false photon candidates. To better constrain this background, the signal region (SR) is split into two parts based on the azimuthal angle $\phi$ of the photon. Energy clusters in the ECAL caused by beam halo muons are observed to concentrate in |sin($\phi$)| $\sim$ 0, whereas photon candidates from all other collision-related processes are uniformly distributed around $\phi$~\cite{Orfanelli:2015}. This motivates the split of the SR into two parts according to $\phi$: vertical SR ($|\phi| >0.5$) and horizontal SR ($|\phi| <0.5$) . 

\begin{figure}[htb!]
    \centering
        \includegraphics[width=1.0\columnwidth]{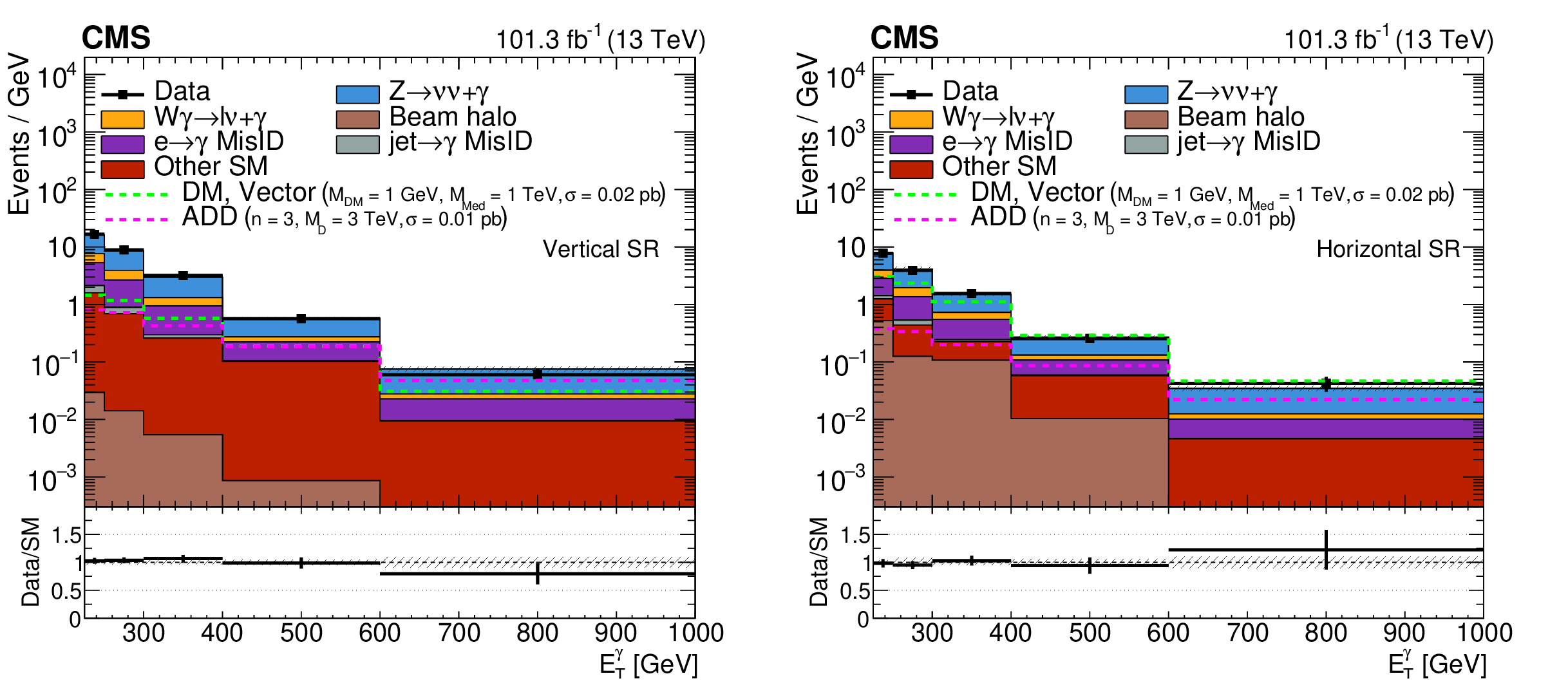}
        \caption{Comparison of data and background post-fit distributions for the vertical (left) and horizontal (right) SRs for the combination of 2017 and 2018 data sets. Templates for signal hypotheses are shown overlaid as light green and magenta dashed lines along with their cross section values. The last bin of the distribution includes the overflow events.}
        \label{fig:monophoton_postfit}
\end{figure}

The main selections include $p_{\mathrm{T}}^\text{miss}$ filters, azimuthal angle ($\phi$) between $p_{\mathrm{T}}^\text{miss}$ direction and photon to be larger than 0.5 radians, $p_{\mathrm{T}}^\text{miss}$ $>200$ GeV, and selections to reject events having electrons, muons and photons. The main uncertainties are related to higher order QCD, electroweak uncertainties, parton distribution function uncertainties and photon identification and scale factor uncertainties. In this analysis too, four control samples (single-electron, double-electron, single-muon, double-muon) are defined for each of eras of data processing by reversing the rejection criteria of electrons and muons. These control samples are used in a global maximum likelihood fit along with the SR to extract the signal. Fig.~\ref{fig:monophoton_postfit} shows the SR postfit plots. No excess of events over the SM background expectation is observed. 

\begin{figure}[htb!]
    \centering
        \includegraphics[width=1.0\columnwidth]{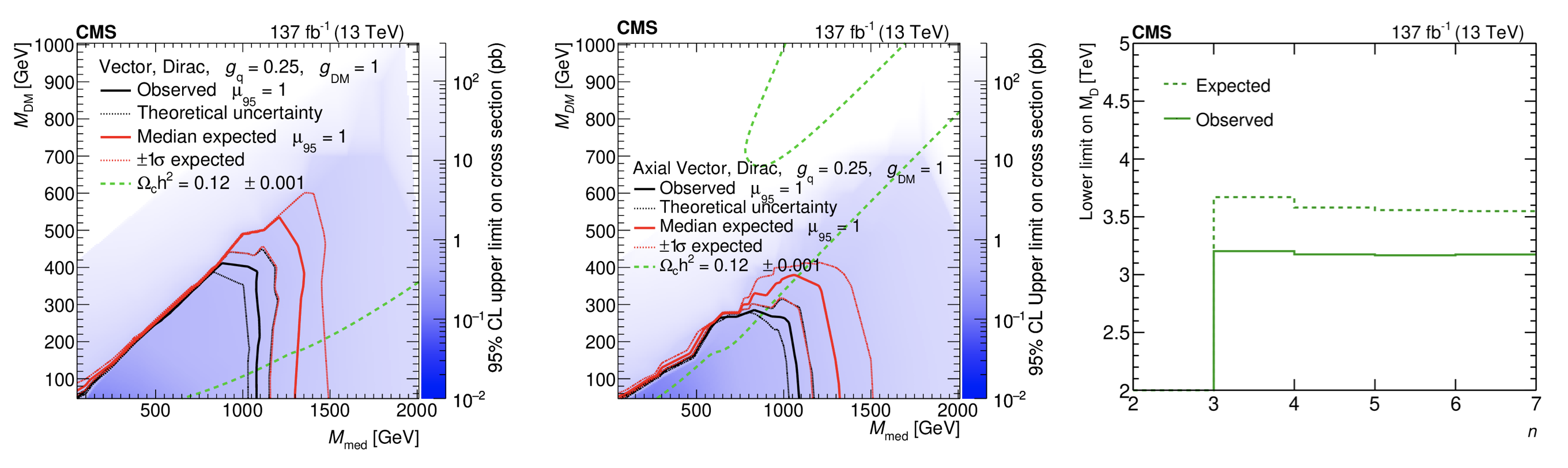}
        \caption{The ratio of 95\% CL upper cross section limits to the theoretical cross section, for simplified DM models with vector (left) and axial-vector (middle) mediators. The region below and to the left of the observed contour is excluded. Lower limit on the fundamental Planck scale MD as a function of the number of extra dimensions n (right). These results are using the full Run 2 dataset.}
        \label{fig:monophoton_limits}
\end{figure}

The analysis which uses only 2017 and 2018 data is combined with a 2016-only version~\cite{monophoton_2016} using a joint likelihood function to enhance the sensitivity. The results are interpreted in terms of DM scenarios and ADD extra spacetime dimensions models as shown in Fig.~\ref{fig:monophoton_limits}. For the simplified dark matter production models considered, the observed (expected) lower limit on the mediator mass is 1085 (1300) GeV in both cases for a 1 GeV dark matter particle mass. While for the model with extra spatial dimensions, values of the effective Planck scale MD up to 3.2 TeV are excluded between 3 and 6 extra dimensions.

\section{Mono-Top Search}
This analysis~\cite{monotop_paper} focuses on the mono-top signature which consists of a single top quark and a large amount of $p_{\mathrm{T}}^\text{miss}$ in the final state (the term top quark denotes both the top quark and the top antiquark).
\noindent\begin{minipage}[t]{0.47\textwidth}
  \vspace{0pt}
   Mono-top production in the SM cannot occur at tree level, but only at one-loop level and at higher orders in quantum chromodynamics (QCD) perturbation theory. Furthermore, Cabibbo suppression~\cite{Cabibo_1,Cabibo_2} as well as the GIM mechanism~\cite{GIM} reduce the cross section for the production of this signature by SM processes. Thus, searches for mono-top production in Beyond-SM theories benefit from low expected SM backgrounds. By extending the SM with the non-resonant mono-top model~\cite{Andrea_2011}, mono-top production becomes possible at tree-level as shown in Fig.~\ref{fig:monotop_feynmandiag}. The dominant SM backgrounds are from Z($\nu\nu$)+jets and W($l\nu$)+jets processes.

   \vspace{0.2cm}

   The main selections include $p_{\mathrm{T}}^\text{miss}$ filters, azimuthal angle ($\phi$) between $p_{\mathrm{T}}^\text{miss}$ direction and jets to be larger than 0.5 radians, $p_{\mathrm{T}}^\text{miss}$ $>350$ GeV, and selections to reject events having electrons, muons, and photons.
\end{minipage}%
\hfill
\begin{minipage}[t]{0.5\textwidth}
  \vspace{0pt}
  \includegraphics[width=0.8\linewidth]{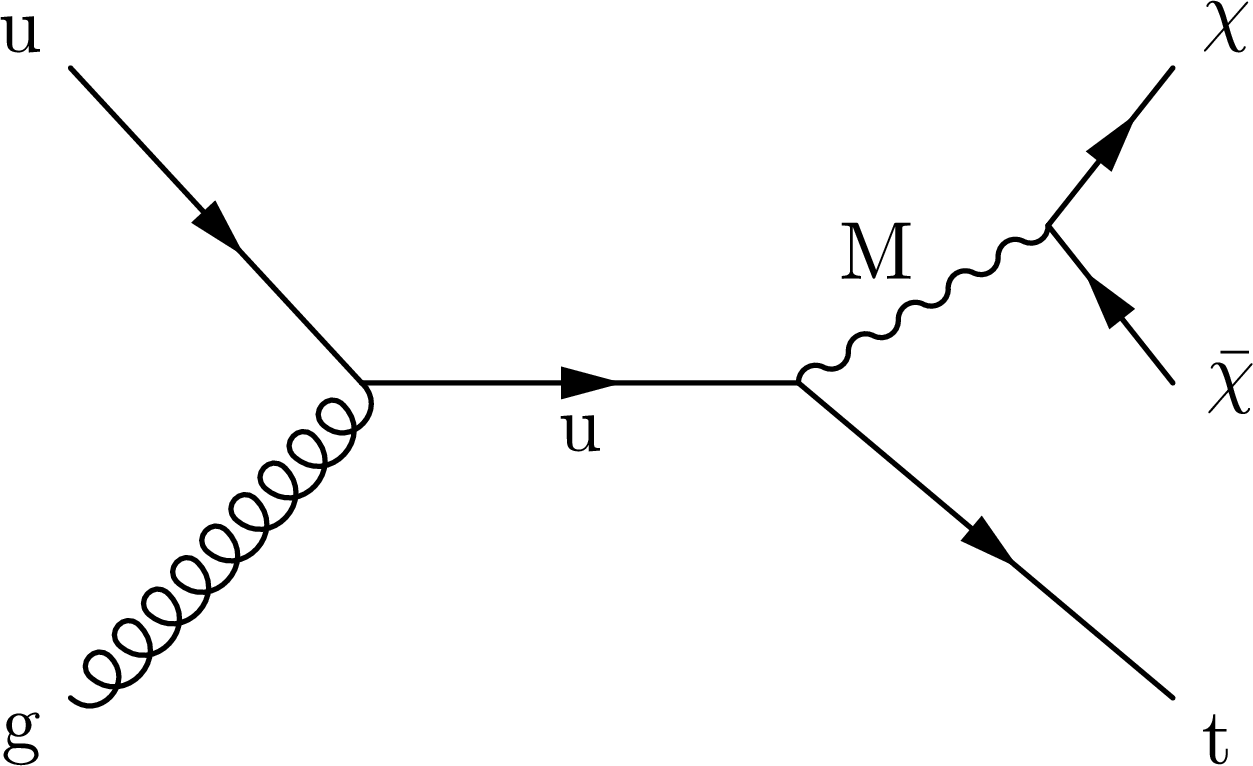}
  \captionof{figure}{Representative Feynman diagram of nonresonant mono-top production at tree level via a flavor-changing neutral current mediated by the spin-1 boson M. The off-shell up quark (u) decays into an on-shell top quark (t) and an M boson. The M boson decays directly to a pair of DM candidates $\chi$ and $\bar{\chi}$.}
  \label{fig:monotop_feynmandiag}
\end{minipage}

\begin{figure}[htb!]
    \centering
        \includegraphics[width=1.0\columnwidth]{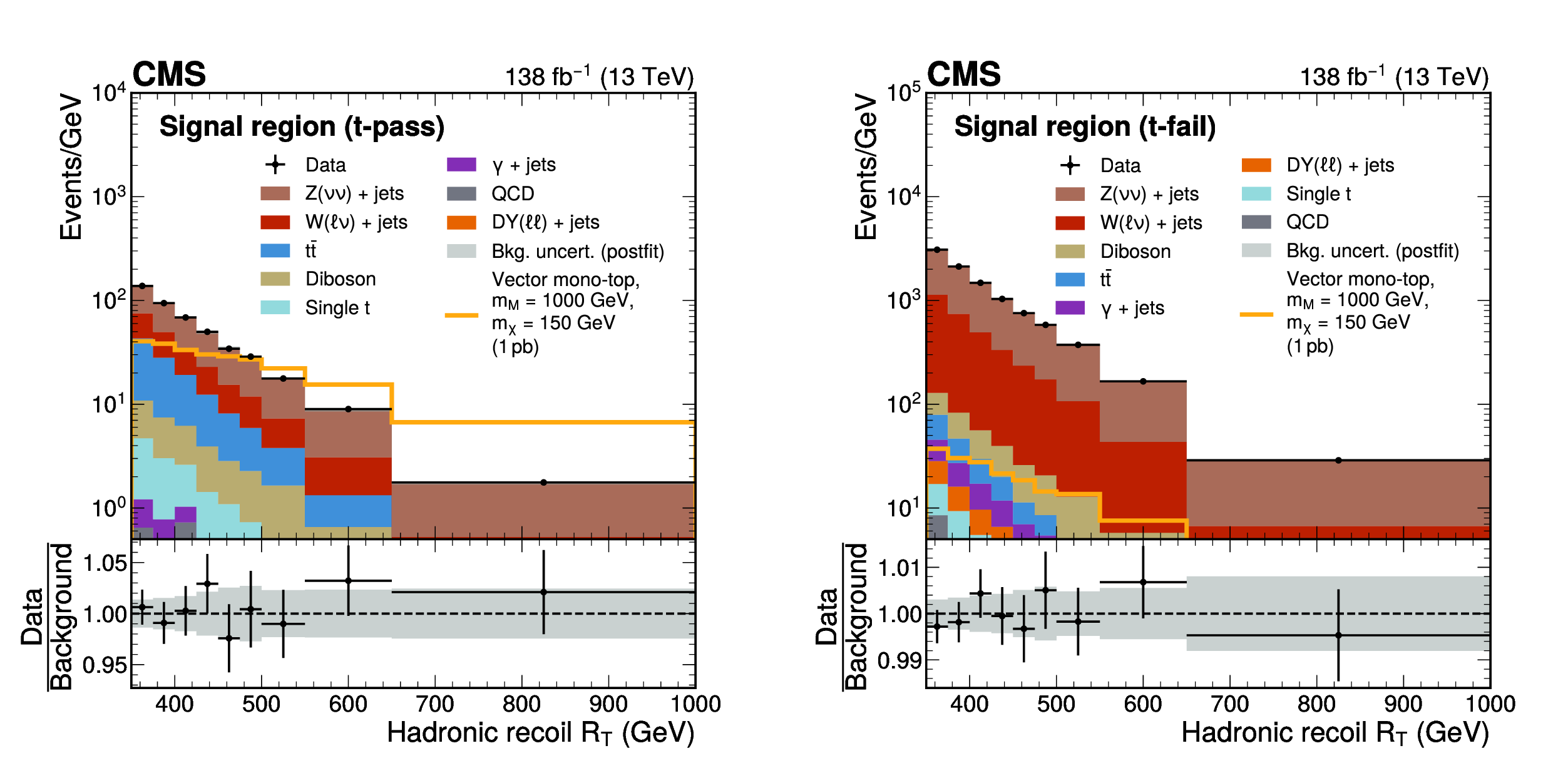}
        \caption{Comparison of data and background post-fit distributions for the signal regions.}
        \label{fig:monotop_postfit}
\end{figure}

 ParticleNet tagger~\cite{Qu_2020} is used to distinguish top jets against QCD jets and also to split the signal region into two categories depending on the tagger score. The main uncertainties are related to higher order QCD, electroweak uncertainties, parton distribution function uncertainties and top-tagging calibration uncertainties. In this analysis too, four control samples (single-electron, double-electron, single-muon, double-muon) are defined for each data-taking era by reversing the rejection criteria of electrons and muons. These control samples are used in a global maximum likelihood fit along with the SR to extract the signal. Fig.~\ref{fig:monotop_postfit} shows the SR postfit plots. No excess of events over the SM background expectation is observed. 

\begin{figure}[htb!]
    \centering
        \includegraphics[width=1.0\columnwidth]{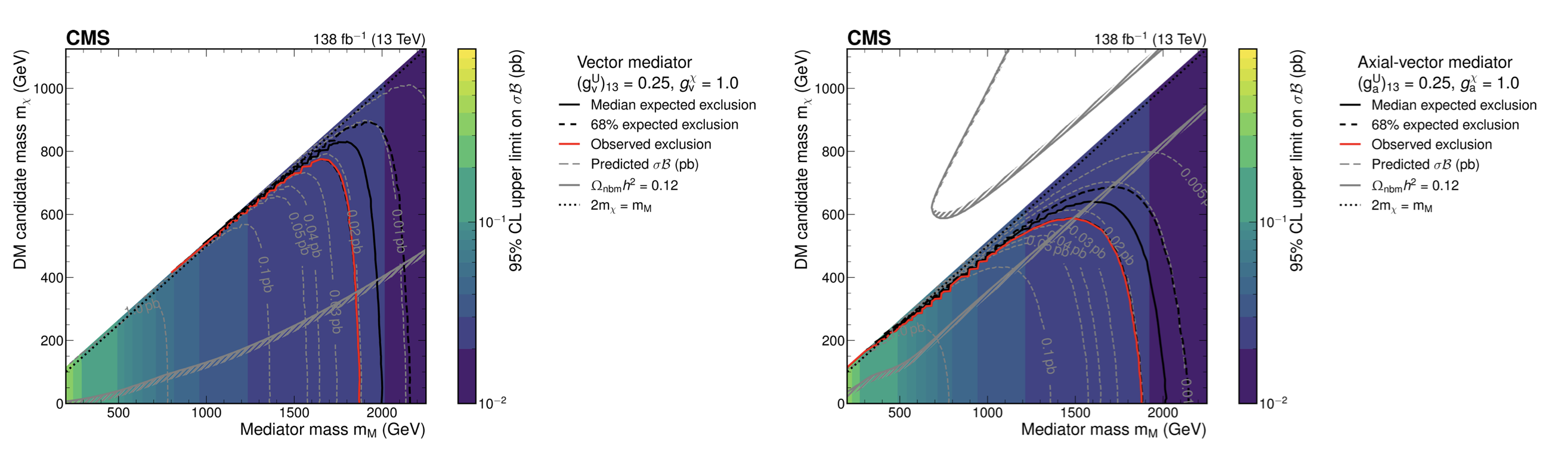}
        \caption{Upper limits at 95\% CL on $\sigma B$ of mono-top production presented in the two-dimensional plane spanned by the mediator and DM candidate masses for a mediator mass between 200 and 2250 GeV and a DM candidate mass between 1 and 1125 GeV only considering on-shell decays of the mediator to the DM candidates. The mediator has vector couplings to quarks and DM candidates in the left plot and axial-vector couplings in the right plot.}
        \label{fig:monotop_limits}
\end{figure}

The results are interpreted in terms of vector and axial-vector mediator DM scenarios as shown in Fig.~\ref{fig:monotop_limits}. In a vector (axial-vector) coupling scenario, masses of the spin-1 mediator are excluded up to 1.85 (1.85) TeV with an expectation of 2.0 (2.0) TeV, whereas masses of the dark matter candidates are excluded up to 0.75 (0.55) TeV with an expectation of 0.85 (0.65) TeV. 

\newpage
\printbibliography

\end{document}